%% file: nh-dirac.tex
\documentclass[aps,twocolumn,prb,superscriptaddress,citeautoscript]{revtex4-2}
\usepackage{amsmath}
\usepackage{amssymb}
\usepackage{graphicx}
\usepackage{color}
\usepackage{bm}
\usepackage{physics}

\begin{document}

\title{Scattering Dynamics and Boundary States of a Non-Hermitian Dirac Equation}

\author{Yun Yong Terh}
\thanks{These authors contributed equally to this work.}
\affiliation{School of Physical and Mathematical Sciences, Nanyang Technological University,
Singapore 637371, Singapore}

\author{Rimi Banerjee}
\thanks{These authors contributed equally to this work.}
\affiliation{School of Physical and Mathematical Sciences, Nanyang Technological University,
Singapore 637371, Singapore}

\author{Haoran Xue}
\affiliation{School of Physical and Mathematical Sciences, Nanyang Technological University,
Singapore 637371, Singapore}


\author{Y. D. Chong}
\email{yidong@ntu.edu.sg}
\affiliation{School of Physical and Mathematical Sciences, Nanyang Technological University, Singapore 637371, Singapore}
\affiliation{Centre for Disruptive Photonic Technologies, Nanyang Technological University, Singapore, 637371, Singapore}

\begin{abstract}
We study a non-Hermitian variant of the (2+1)-dimensional Dirac wave equation that hosts a real energy spectrum with pairwise-orthogonal eigenstates.  In the spatially uniform case, the Hamiltonian's non-Hermitian symmetries allow its eigenstates to be mapped to a pair of Hermitian Dirac subsystems.  When a wave is transmitted across an interface between two spatially uniform domains with different parameters, an anomalous form of Klein tunneling occurs, whereby reflection is suppressed while the transmittance is substantially higher or lower than unity.  The interface can even function as a simultaneous laser and coherent perfect absorber.  Remarkably, the violation of flux conservation occurs entirely at the interface, with no wave amplification or damping taking place in the bulk.  Moreover, at energies within the Dirac mass gaps, the interface can support exponentially localized boundary states with real energies.  These features of the continuum model can be reproduced in non-Hermitian lattice models.
\end{abstract}

\maketitle

\section{Introduction}

The Dirac equation, formulated by Dirac in 1929 to describe relativistic electrons and positrons \cite{Dirac1928}, has long fascinated physicists with its rich implications, from the determination of the electron's magnetic moment to the phenomenon of Klein tunneling \cite{Klein1929}.  In recent years, the Dirac equation has also taken on a key role in condensed matter and other non-relativistic systems: 2D and 3D Dirac equations are used to describe electronic states in materials such as graphene \cite{Novoselov2005, Zhang2005, CastroNeto2009, young2012DiracSemimetal}, and bandstructures near a topological phase transition typically simplify to a Dirac equation or one of its relatives \cite{haldane1988model, hasan2010colloquium}.  Moreover, by exploiting the mathematical similarity of single-particle Schr\"{o}dinger equations and classical wave equations, the Dirac equation can be realized in a variety of classical wave metamaterials \cite{RevModPhys.91.015006, XiangYves}, such as photonic and acoustic crystals \cite{Sepkhanov2007, Zhang2008, Ochiai2009, Bittner2010, Huang2011, Zhong2011, Torrent2013, Lu2014}, resonator arrays \cite{Yang2017}, and waveguide arrays \cite{Tran2014, Koke2020, Leykam2016}.

The study of classical wave metamaterials has also motivated interest in non-Hermitian wave equations, due to the ease with which gain and/or loss can be controlled in such systems.  The first non-Hermitian Hamiltonians to be thus studied were those obeying parity/time-reversal (PT) symmetry, a non-Hermitian symmetry originally conceived by Bender and Boettcher for a hypothetical generalization of quantum mechanics \cite{bender1998real, bender1999pt}.  PT symmetric Hamiltonians have now been realized in a wide range of experimental platforms, such as optical waveguide arrays \cite{ElGanainy07, ruter2010observation, el2018non}.  Their two most remarkable features---the ability to exhibit real energy spectra despite being non-Hermitian, and the spontaneous breaking of PT symmetry whereby pairs of real energies merge and turn complex---have been found to produce many interesting applications, such as unidirectional invisibility \cite{lin2011unidirectional} and laser mode stabilization \cite{zhao2018topological}.  More recently, there has also been progress on other aspects of non-Hermitian physics, including the topology of non-Hermitian bandstructures \cite{rudner2009topological, liang2013topological, shen2018topological, bergholtz2021exceptional}, which is associated with phenomena such as non-Hermitian topological modes \cite{Liang2013, Lee2016, Leykam2017} and the non-Hermitian skin effect \cite{HN1996, Martinez2018, Slager2020, Okuma2020}.

Recently, Xue \textit{et al.}~proposed a class of non-Hermitian Hamiltonians whose bandstructures host non-Hermitian Dirac cones \cite{xue2020non}.  These Hamiltonians obey a set of non-Hermitian symmetries related to but distinct from PT symmetry, leading to the existence of pairs of real energy eigenvalues with eigenvectors that are mutually orthogonal, according to the usual inner product employed for Hermitian systems (by contrast, the eigenvectors of PT symmetric Hamiltonians need not be orthogonal in this sense).  When two such eigenvalues become degenerate, they form a diabolic point (e.g., a Dirac point), unlike the exceptional point degeneracies more commonly found in non-Hermitian systems \cite{el2018non, bergholtz2021exceptional}.  The non-Hermitian Dirac points were shown to have properties matching Dirac points in Hermitian bandstructures, such as the ability to undergo band inversions giving rise to topological boundary states \cite{xue2020non}.  However, these features were explored only in the context of discrete (tight-binding) models.  The nature of the non-Hermitian Dirac equation, which ought to govern the non-Hermitian Dirac-like bands in the continuum limit, was not discussed in detail.

In this paper, we develop the non-Hermitian Dirac equation (NHDE) that governs the non-Hermitian Dirac cones described in Ref.~\onlinecite{xue2020non}, and study its physical implications.  The NHDE takes the form of a 4-component Dirac-like equation in (2+1)D, with a non-Hermitian mass operator.  By a variation of Dirac's argument \cite{Dirac1928}, we show that it supports relativistic (hyperbolic) dispersion relations.  The NHDE can be transformed into a pair of 2-component Hermitian Dirac subsystems, whose Dirac masses correspond to the eigenvalues of the non-Hermitian mass operator.  Notably, the mass gaps depend on a non-Hermitian parameter corresponding to on-site gain and loss, so it is possible for band inversions to occur via gain/loss tuning \cite{TakataNotomi2018, Kawabata2018, Luo2019, Wu2020, xue2020non}.

Next, we study the nature of wave propagation based on the NHDE.  For a plane wave incident on an interface separating two uniform semi-infinite domains with unequal scalar potential and/or certain choices of Hamiltonian parameters, we show that the NHDE can exhibit exactly the same reflection and transmission characteristics as the Hermitian Dirac equation, including Klein tunneling (i.e., 100\% transmission when the incident energy is far detuned from the mass gap \cite{Klein1929}).

For other parameter choices, flux conservation is violated, resulting in an anomalous form of Klein tunneling whereby reflection is suppressed but the transmitted flux is larger or smaller than the incident flux.  Remarkably, the violation of flux conservation  appears to originate at the domain wall itself, since wave amplification or damping does not occur in the bulk domains, which host real-energy plane wave solutions.  At certain energies, the violation of flux conservation can become so extreme that the domain wall acts as a simultaneous laser and coherent perfect absorber \cite{chong2010, baranov2017, longhi2010cpalaser, chong2011pt}, spontaneously emitting flux with no input while also perfectly absorbing a specific configuration of input waves.  On the other hand, at energies lying in the mass gaps, the domain wall can host chiral boundary states that behave much like the topological boundary states of the Hermitian Dirac equation.  These boundary states have real energies, even if the domain wall violates flux conservation at energies outside the mass gaps.

\section{Non-Hermitian 2D Dirac Equation}
\label{sec:basics}

Consider the following time-dependent Dirac-like equation, defined in 2+1 dimensions:
\begin{equation}
  i\frac{\partial \psi}{\partial t} =
  \Big(\mathcal{M} - i\alpha_1 \partial_1 - i\alpha_2\partial_2\Big)
  \psi(\mathbf{r},t).
  \label{diracform}
\end{equation}
Here, $\mathbf{r} = (x,y)$ denotes 2D spatial coordinates, $t$ is the time coordinate, $\partial_i = \partial/\partial x_i$ for $i\in \{1,2\}$, and $\{\mathcal{M}, \alpha_{1}, \alpha_2\}$ are matrices to be specified.  If we choose $\mathcal{M} = m\sigma_3$ and $\alpha_i = \sigma_i$, where $\{\sigma_i\}$ are the Pauli matrices, Eq.~\eqref{diracform} reduces to the standard (2+1)D Dirac equation, with $\psi$ a two-component wavefunction, $m$ representing the Dirac mass, and the speed of light and $\hbar$ both set to unity.  We are interested in a \textit{non-Hermitian} Dirac equation obtained with an alternative choice of $\mathcal{M}$ and $\alpha_{1,2}$.  We shall see that the non-Hermitian model can exhibit numerous features normally associated with the Hermitian Dirac equation \cite{xue2020non}.

Before proceeding, we define
\begin{align}
    \Sigma_\mu = \begin{bmatrix}
    0 &\sigma_\mu \\
    \sigma_\mu &0
    \end{bmatrix} = \tau_1 \sigma_\mu,
    \label{sigmamatrices}
\end{align}
where $\mu\in\{0,1,2,3\}$, and $\tau_\nu\sigma_\mu$ denotes the tensor product between Pauli matrices $\tau_\nu$ and $\sigma_\mu$.  In Ref.~\onlinecite{xue2020non}, it was shown that if a Hamiltonian $H$ obeys the symmetries
\begin{align}
    \Sigma_0 H \Sigma_0 &= H^\dagger, \label{sym1}\\
    \{H, \Sigma_3\Sigma_1 T\} &= 0,  \label{sym2}
\end{align}
where $T$ is the complex conjugation operator, then there are two consequences.  First, the eigenvalues are either real (symmetry-unbroken) or appear in complex conjugate pairs (symmetry-unbroken).  Second, if $|\psi\rangle$ is an eigenstate with eigenvalue $E$, then $|\psi'\rangle = \Sigma_1 \Sigma_3 T |\psi\rangle$ is an orthogonal eigenstate with energy $-E^*$ (where $T$ is the complex conjugation operator and the orthogonality relation $\langle\psi'|\psi\rangle = 0$ uses the standard Hermitian inner product).  Thus, the non-Hermitian $H$ retains a remnant of the two key properties of Hermitian operators, eigenvalue reality and eigenstate orthogonality.  By contrast, PT symmetric non-Hermitian Hamiltonians can exhibit real eigenvalues, but generally have non-orthogonal eigenstates \cite{bender1998real, bender1999pt, bender2007making}.

\begin{figure}
  \centering\includegraphics[width=0.49\textwidth]{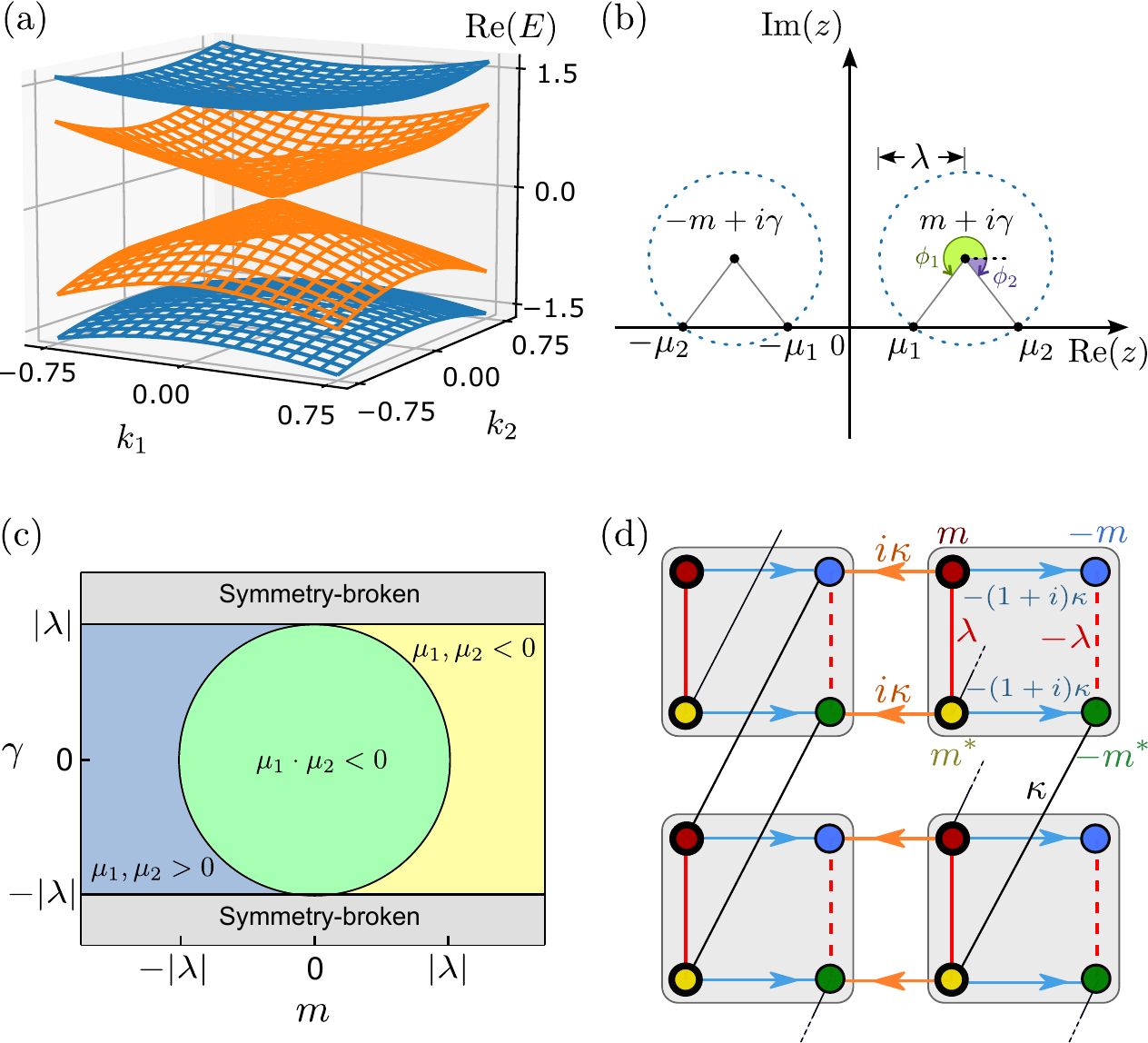}
  \caption{(a) Bandstructure of the 2D non-Hermitian Dirac equation (NHDE).  All the eigenvalues are real as the semi-Hermiticity symmetry is unbroken.  The parameters are $\lambda=0.8$, $m=0.6$, and $\gamma = \sqrt{\lambda^2 - m^2}$, chosen so that one of the Dirac masses vanishes. (b) Geometric relationship between the model parameters $\{\lambda, m,\gamma\}$ and the Dirac masses $\mu_{1,2}$, based on Eqs.~\eqref{eq:phin}--\eqref{eq:cosphin}.  (c) Phase diagram of the 2D NHDE.  The symmetry-unbroken regime $|\gamma| < |\lambda|$ consists of three distinct phases, determined by the signs of $\mu_{1,2}$.  (d) A 2D non-Hermitian lattice that yields the NHDE in the long-wavelength limit.  The unit cells (gray boxes) each contain 4 sites, and every hopping is Hermitian.  Reciprocal hoppings are indicated by solid and dashed lines, while nonreciprocal hoppings are indicated by arrows and labelled by the hopping amplitude in the arrow direction.}
  \label{fig:model}
\end{figure}

For brevity, we will refer to Hamiltonians obeying Eqs.~\eqref{sym1}--\eqref{sym2} as ``semi-Hermitian''.  When semi-Hermicity is unbroken, the eigenvalues form two pairs, $\{ \pm E_1 \}$ and $\{\pm E_2\}$. Within a pair (but not between pairs), the corresponding eigenstates are orthogonal.

Returning to the 2D Dirac-type equation \eqref{diracform}, let $\mathcal{M}$ be the $4\times4$ matrix
\begin{equation}
  \mathcal{M} = \begin{bmatrix}
    W &V\\
    V &W^\dagger
  \end{bmatrix}, \label{mdef}
\end{equation}
where
\begin{align}
  W &= \begin{bmatrix}
    m + i \gamma &b\\
    b^* & -m +i\gamma
  \end{bmatrix}, \label{wdef} \\
  V &= V^\dagger = \begin{bmatrix}
        \lambda &c\\
        c^* &-\lambda
  \end{bmatrix},
\end{align}
for $b, c \in \mathbb{C}$ and $m, \gamma, \lambda \in \mathbb{R}$.  Then $\mathcal{M}$ obeys Eq.~\eqref{sym1}--\eqref{sym2} and is semi-Hermitian.  We will see later how $\mathcal{M}$ could be physically realized.  By analogy with the Hermitian Dirac equation, we seek $\alpha_1, \alpha_2$ such that 
\begin{align}
    \{\alpha_i, \alpha_j\}& = 2\delta_{ij}, \quad i,j\in\{1,2\}, \label{eq:alphaanticom}\\
    \{\mathcal{M}, \alpha_i\}& = 0. \label{eq:alpha_Manticom}
\end{align}
We do not assume $\mathcal{M}^2$ to be proportional to the identity, in a departure from the formulation of the Hermitian Dirac equation.  If we take the representation $\alpha_i = \Sigma_i$, which satisfies Eq.~\eqref{eq:alphaanticom}, then Eq.~\eqref{eq:alpha_Manticom} implies $b = c = 0$, and $\mathcal{M}$ has the form
\begin{align}
  \begin{aligned}
    \mathcal{M}
    &= m \sigma_3 + i\gamma \tau_3 + \lambda \tau_1 \sigma_3.
  \end{aligned}
  \label{eq:Mmatrixcompact}
\end{align}
We thus arrive at the non-Hermitian Dirac equation (NHDE)
\begin{align}
  \begin{aligned}
  i\frac{\partial \psi}{\partial t} &=
  \Big(\mathcal{M} - i\Sigma_1 \partial_1 - i\Sigma_2\partial_2\Big)
  \psi(\mathbf{r},t)\\
  &= \tau_1 \left(- i \sum_{j=1}^2 \sigma_j \partial_j + \lambda \sigma_3\right)
  + m \sigma_3 + i\gamma \tau_3.
  \end{aligned}
  \label{diracform2}
\end{align}
Thus, the model can also be interpreted as a pair of Dirac ``valleys'' that interact via both a Hermitian coupling, $m\sigma_3$, and a non-Hermitian coupling, $i\gamma\tau_3$.

To derive the dispersion relation, recall that in the standard Dirac equation, one takes the square of the Hamiltonian and uses the anticommutivity of the Dirac matrices \cite{Dirac1928}.  A variant of this procedure can be applied to the non-Hermitian equation \eqref{diracform2}.  Let $|\psi\rangle$ be an eigenstate of energy $E$ and momentum $\mathbf{k}$. Squaring the Hamiltonian and using Eqs.~\eqref{eq:alphaanticom}---\eqref{eq:alpha_Manticom}, we obtain
\begin{equation}
  \Big(\mathcal{M}^2 + |\mathbf{k}|^2\Big) \ket{\psi}
  = E^2 \ket{\psi},
\end{equation}
where $\mathcal{M}^2 \ne I$.  This implies that $\ket{\psi}$ is also an eigenvector of $\mathcal{M}^2$.  If the semi-Hermiticity of $\mathcal{M}$ is unbroken, it has real eigenvalues $\pm \mu_n$, where $n = 1,2$; then $\mathcal{M}^2$ has doubly degenerate eigenvalues $\mu_n^2$, and
\begin{align}
  E = \pm \sqrt{\mu_n^2 + |\mathbf{k}|^2}. \label{eq:2Dlineardispersion}
\end{align}
Therefore, for unbroken semi-Hermiticity, the dispersion relation consists of a pair of relativistic hyperbolas, as shown in Fig.~\ref{fig:model}(a).

Note that even if $\ket{\psi}$ is an eigenvector of $\mathcal{M}^2$, it need not be an eigenvector of $\mathcal{M}$; in general, $\ket{\psi}$ could be a superposition of two eigenvectors of $\mathcal{M}$ whose eigenvalues have the same square.  To find these, take the ansatz $|\varphi\rangle|\!\updownarrow\rangle$, where $\sigma_3 |\!\updownarrow\rangle = \pm |\!\updownarrow\rangle$.  Applying Eq.~\eqref{eq:Mmatrixcompact}:
\begin{align}
  \mathcal{M} \, |\varphi^n_\pm\rangle\, |\!\updownarrow\rangle
  &= \pm \mu_n |\varphi^n_\pm\rangle \, |\!\updownarrow\rangle, \label{Meq0} \\
  H_\pm |\varphi_\pm^n\rangle &= \pm \mu_n |\varphi_\pm^n\rangle, \label{Hpmeigenproblem}\\
  H_\pm &= \pm \left(\lambda \tau_1 + m \pm i\gamma \tau_3 \right),
  \label{Hplusminus}
\end{align}
where $H_\pm$ is a $2\times 2$ matrix, and $n$ indexes its eigenvectors.  These sub-Hamiltonians obey the PT symmetry \cite{bender1998real, bender1999pt, bender2007making}
\begin{equation}
  [H_\pm,\tau_1T] = 0,
\end{equation}
and are related by
\begin{equation}
  H_\pm = - \tau_1 H_\mp\tau_1. \label{Hplusminus_relation}
\end{equation}
Let us take
\begin{align}
  \ket{\varphi^n_+} = \frac{1}{\sqrt{2}} \begin{bmatrix}
    1\\e^{i\phi_n}
  \end{bmatrix}, \quad
  |\varphi_\pm^n\rangle = \tau_1 |\varphi_\mp^n\rangle,
  \label{eq:varphiform}
\end{align}
for some $\phi_n$.  Using Eqs.~\eqref{Hpmeigenproblem}--\eqref{Hplusminus}, we find that
\begin{equation}
  \mu_n = m+i\gamma + \lambda e^{i\phi_n}.
\end{equation}
The real and imaginary parts of this equation are
\begin{align}
  \cos\phi_n &= (\mu_n - m)/\lambda, \label{eq:cosphin} \\
  \sin\phi_n &= - \gamma/\lambda. \label{eq:phin}
\end{align}

Given $m$ and $\gamma$, Eqs.~\eqref{eq:cosphin}--\eqref{eq:phin} have two sets of solutions, $\{\phi_1, \mu_1\}$ and $\{\phi_2, \mu_2\}$.  Fig.~\ref{fig:model}(b) illustrates the geometric relationship between these quantities.
The above derivation also clarifies how the regime of unbroken semi-Hermiticity, introduced in Ref.~\onlinecite{xue2020non}, corresponds to the PT-unbroken phase for an underlying pair of  $2\times2$ sub-Hamiltonians \eqref{Hplusminus}.

Now, for fixed $n \in \{1,2\}$, consider
\begin{align}
  \ket{\psi_n}= c^n_+ \ket{\varphi_+^n} \ket{\uparrow} + c^n_- \ket{\varphi_-^n} \ket{\downarrow}.
\end{align}
Plugging this into the NHDE, and using the above properties of $\ket{\varphi_\pm^n}\ket{\updownarrow}$, we obtain
\begin{align}
  \Big( k_1 \sigma_1 + k_2 \sigma_2 + \mu_n \sigma_3 \Big) \begin{bmatrix}
    c_+^n \\ c_-^n \end{bmatrix} = E \begin{bmatrix}
    c_+^n \\ c_-^n \end{bmatrix},
  \label{2dherm}
\end{align}
which is the Hermitian 2D Dirac equation with mass $\mu_n$.

There is one Hermitian 2D Dirac equation \eqref{2dherm} for each $n \in \{1, 2\}$, where $n$ enumerates the solutions to the underlying $2\times 2$ eigenproblem \eqref{Hpmeigenproblem}.  The relationship between the two $2\times2$ sub-Hamiltonians, Eq.~\eqref{Hplusminus_relation}, plays a role similar to particle-hole symmetry.  In the Hermitian limit $\gamma = 0$, this becomes a standard particle-hole symmetry, and the Dirac masses are $\mu_{1,2} = m \pm \lambda$.

We can use the signs of the Dirac masses to derive a phase diagram for the non-Hermitian Dirac Hamiltonian.  This is shown in Fig.~\ref{fig:model}(c), plotted against $m$ and $\gamma$.  If semi-Hermicity is unbroken, there are three distinct phases: inside the circle $\sqrt{m^2+\gamma^2} = |\lambda|$, the Dirac masses have opposite signs, and outside it they are both either positive or negative.  Along the phase boundary, at least one of the masses vanishes.  This occurs when, in Fig.~\ref{fig:model}(b), the dashed blue circle, centered at $m+i\gamma$ and of radius $\lambda$, crosses the origin.

Notably, we can drive the system across the phase boundary by varying the non-Hermitian parameter $\gamma$, while keeping $m$ and $\lambda$ unchanged.  Gain/loss-induced topological phase transitions have also recently been observed in other models \cite{TakataNotomi2018, Kawabata2018, Luo2019, Wu2020, xue2020non}.  These band inversions are associated with boundary states, as discussed in Sec.~\ref{sec:boundstates}.

\section{Non-Hermitian Lattice Model}
\label{sec:2Dlattice}

In this section, we present an exemplary tight-binding lattice model that gives rise to the NHDE, Eq.~\eqref{diracform2}, in its long-wavelength limit.  Such lattices can be realized physically using metamaterials, such as photonic or acoustic structures \cite{RevModPhys.91.015006, OtaTakataOzawaAmo, PartoLiu, Price_2022, Ma2019, Xue2022}.  Previously, in Ref.~\onlinecite{xue2020non}, different tight-binding models satisfying the semi-Hermitian symmetries \eqref{sym1}--\eqref{sym2} have been derived, and it was shown that they can serve as non-Hermitian analogues for Chern insulators, Weyl semimetals, etc.  However, the relationship to the Dirac Hamiltonian was not studied in detail.

Consider the 2D square lattice shown in Fig.~\ref{fig:model}(d).  Each unit cell contains 4 sites with complex on-site potentials.  All inter-site couplings are Hermitian; some, drawn as solid lines and dashes, are reciprocal (i.e., the forward and backward hopping amplitudes are the same), while others are nonreciprocal (i.e., the reverse hopping amplitudes, opposite to the arrows, are complex conjugates of the forward amplitudes).  Such nonreciprocal hoppings are implementable in metamaterials, e.g., by using coupling resonators with appropriate phase shifts \cite{Hafezi2011, Hafezi2013}.


Let $\psi_n$ denote the set of four amplitudes on unit cell $n$. The time-dependent Schr\"{o}dinger equation is
\begin{align}
  i\frac{\partial \psi_n}{\partial t} = \mathcal{H}_1 \psi_n +
  \sum_{i=1,2}\left( \mathcal{J}_{i} \, \psi_{n-r_i}
  + \mathcal{J}_i^\dagger \, \psi_{n+r_i} \right),
\end{align}
where $\mathcal{H}_1$ is the Hamiltonian for a single unit cell, and $n\pm r_i$ denotes the unit cell displaced by one period forward/backward in the $i$-th direction by one lattice constant $\ell$.  The $\mathcal{J}_i$ matrices are
\begin{equation}
    \mathcal{J}_1 = \begin{bmatrix}
    0 &0 &0 &0\\
    0 &0 &i\kappa &0\\
    0 &0 &0 &0\\
    i\kappa &0 &0 &0
    \end{bmatrix}, \;\;
    \mathcal{J}_2 = \begin{bmatrix}
    0 &0 &0 &\kappa\\
    0 &0 &0 &0\\
    0 &\kappa &0 &0\\
    0 &0 &0 &0
    \end{bmatrix},
    \label{eq:jmatrices}
\end{equation}
where $\kappa$ is the hopping parameter defined in Fig.~\ref{fig:model}(d).  In the long-wavelength (continuum) limit,
\begin{equation}
  \psi_{n\pm r_i} = \psi_n \pm \ell \big(\partial_i \psi \big)_n
  + \frac{\ell^2}{2}\left( \partial^2_i \psi \right)_n + \cdots
\end{equation}
Thus, the time-dependent Schr\"{o}dinger equation becomes
\begin{align}
  \begin{aligned}
  i\frac{\partial \psi_n}{\partial t} &= \left[ \mathcal{H}_1 + \sum_{i=1,2} (\mathcal{J}_i + \mathcal{J}_i^\dagger) \right] \psi_n \\
  &+ \sum_{i=1,2}(\mathcal{J}_i^\dagger - \mathcal{J}_i)\; \ell\;
  \big(\partial_i\psi\big)_n + \mathcal{O}(\ell^2),
  \end{aligned}
\end{align}
where $\mathcal{H}_1$ is the Hamiltonian for one unit cell,
\begin{align}
    \mathcal{H}_1 = \begin{bmatrix}
    m+i\gamma &0 &\lambda &-(1-i)\kappa\\
    0 &-m+i\gamma &-(1+i)\kappa &-\lambda\\
    \lambda &-(1-i)\kappa &m-i\gamma &0\\
    -(1+i)\kappa &-\lambda &0 &-m-i\gamma
    \end{bmatrix}. \nonumber
\end{align}
From Eq.~\eqref{eq:jmatrices}, $\mathcal{J}_i^\dagger - \mathcal{J}_i = -i\kappa\Sigma_i$.  Omitting terms of $\mathcal{O}(\ell^2)$ and higher, we arrive at a wave equation $i\partial \psi/\partial t = \mathbf{H}\psi(\mathbf{x},t)$.  The effective Hamiltonian is
\begin{equation}
  \mathbf{H} = \mathcal{M} -i \kappa \ell \left(\Sigma_1\, \partial_1 + \Sigma_2\, \partial_2\right), \\
\label{eq:H2D}
\end{equation}
where $\mathcal{M}$ has precisely the form given in Eq.~\eqref{eq:Mmatrixcompact}.  After taking the normalization $\kappa\ell = 1$, this is our desired non-Hermitian 2D Dirac equation.

\begin{figure}
  \centering\includegraphics[width=0.49\textwidth]{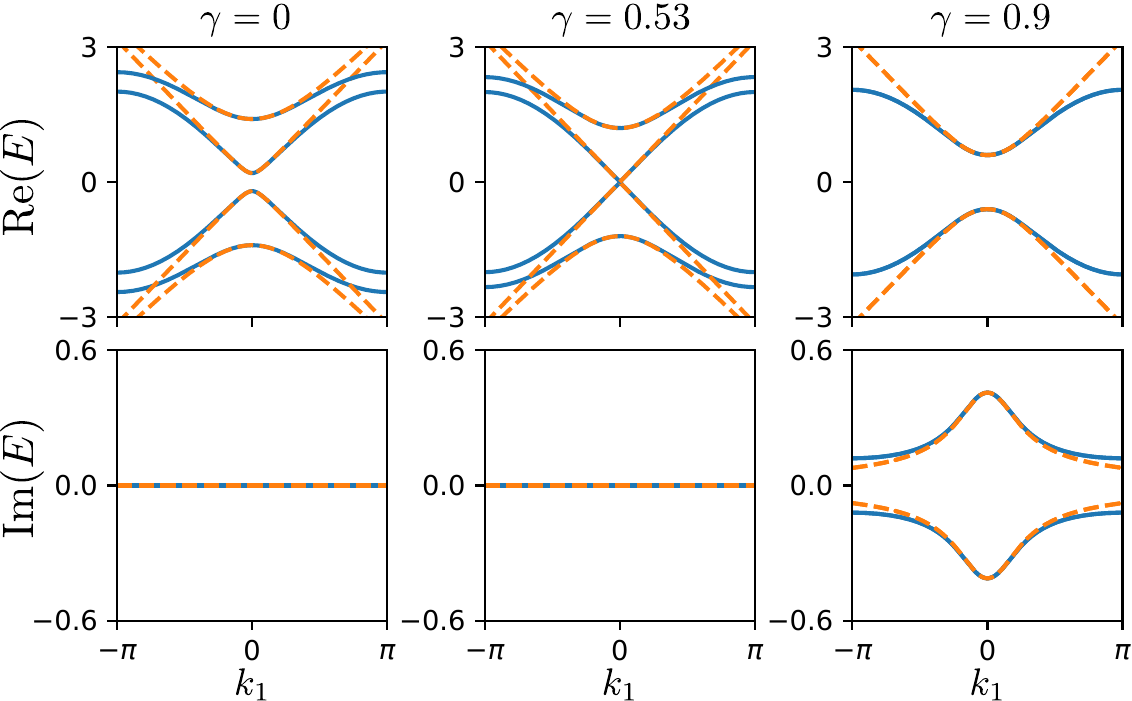}
  \caption{Complex bandstructures of the lattice (solid lines) and the continuum NHDE (dashes), plotted against $k_1$ for $k_2 = 0$ and various choices of $\gamma$.  The other parameters are $m=0.6$, $\lambda=0.8$, and $\kappa=1$. }
  \label{fig:dispersion}
\end{figure}

The lattice's momentum space Hamiltonian  is
\begin{align}
    H(\mathbf{k}) = \begin{bmatrix}
    m &0 &\lambda &c_{\mathbf{k}}\\
    0 &-m^* & c_{\mathbf{k}}^* &-\lambda\\
    \lambda &c_{\mathbf{k}} &m^* &0\\
    c_{\mathbf{k}}^* &-\lambda &0 &-m
    \end{bmatrix},
\end{align}
where
\begin{equation}
  c_{\mathbf{k}} = \kappa[-i(e^{ik_1\ell}-1) + (e^{-ik_2\ell} -1)].
\end{equation}

In Fig.~\ref{fig:dispersion}(a), we plot the complex bandstructure versus $k_1$ with fixed $k_2 = 0$.  The bands obtained from the lattice model are plotted as solid blue lines, while the dispersion relations for the continuum model are plotted as orange dashes.  Results are shown for three different values of $\gamma$, with $m$, $\lambda$, and $\kappa$ fixed; this corresponds to traversing the phase diagram in Fig.~\ref{fig:model}(c) along a horizontal line, passing through a phase boundary.  For small quasimomenta, the band energies of the lattice model approximate the hyperbolic dispersion relations of the continuum model, as expected.  For $\gamma = \pm \sqrt{\lambda^2 - m^2}$, one of the Dirac masses vanishes as predicted by \ref{fig:model}(c).  This gap-closing can be observed in both the lattice and continuum models.

This bandstructure also hosts a second set of non-Hermitian Dirac cones at $\mathbf{k}_{\mathrm{M}} = {(\pi/2,\pi/2)}$.  When one of the Dirac masses vanishes at $\mathbf{k}_\Gamma = (0, 0)$, as described in the previous paragraph, the band gap also closes at $\mathbf{k}_{\mathrm{M}}$, similar to valley Hall systems \cite{haldane1988model}.

For the 1D version of the NHDE, we can use a dimensionally reduced version of the lattice model with only real hoppings, as described in Sec.~\ref{sec:klein}.

\section{Scattering from a domain wall}
\label{sec:klein}

We now ask how the NHDE is affected by spatial non-uniformity.  As discussed in Sec.~\ref{sec:basics}, the spatially uniform model with unbroken semi-Hermiticity behaves as a pair of Hermitian Dirac subsystems [see Eq.~\eqref{eq:varphiform}].  If we now let the parameters vary in space, the Hermitian-like behavior should be expected to persist if the variations keep $\gamma/\lambda$ constant; on the other hand, if $\gamma/\lambda$ is not constant, non-Hermitian effects should become important.  We will focus on two important cases of spatial non-uniformity.  This section will cover reflection and transmission between two spatially uniform domains; the case of boundary states localized at domain walls is studied in Sec.~\ref{sec:boundstates}.

We are particularly interested in the phenomenon of Klein tunneling \cite{Klein1929}, whereby a Dirac particle can transmit perfectly through a potential barrier whose height $\Phi_0$ exceeds its own energy $E$.  (By contrast, for a nonrelativistic particle the transmission is exponentially suppressed.)  Originally predicted for relativistic electrons \cite{Klein1929}, Klein tunneling has also been observed with Dirac Hamiltonians arising in 2D materials \cite{KatsnelsonNovoselov2006, Beenakker2008, StanderHuard2009, YoungKim2009} and photonic lattices \cite{TreidelPeleg2012, DreisowKeil2012, JiangShi2020, OzawaAmo2017, NguyenTran2020}.

\begin{figure*}
\centering
\includegraphics[width=0.95\textwidth]{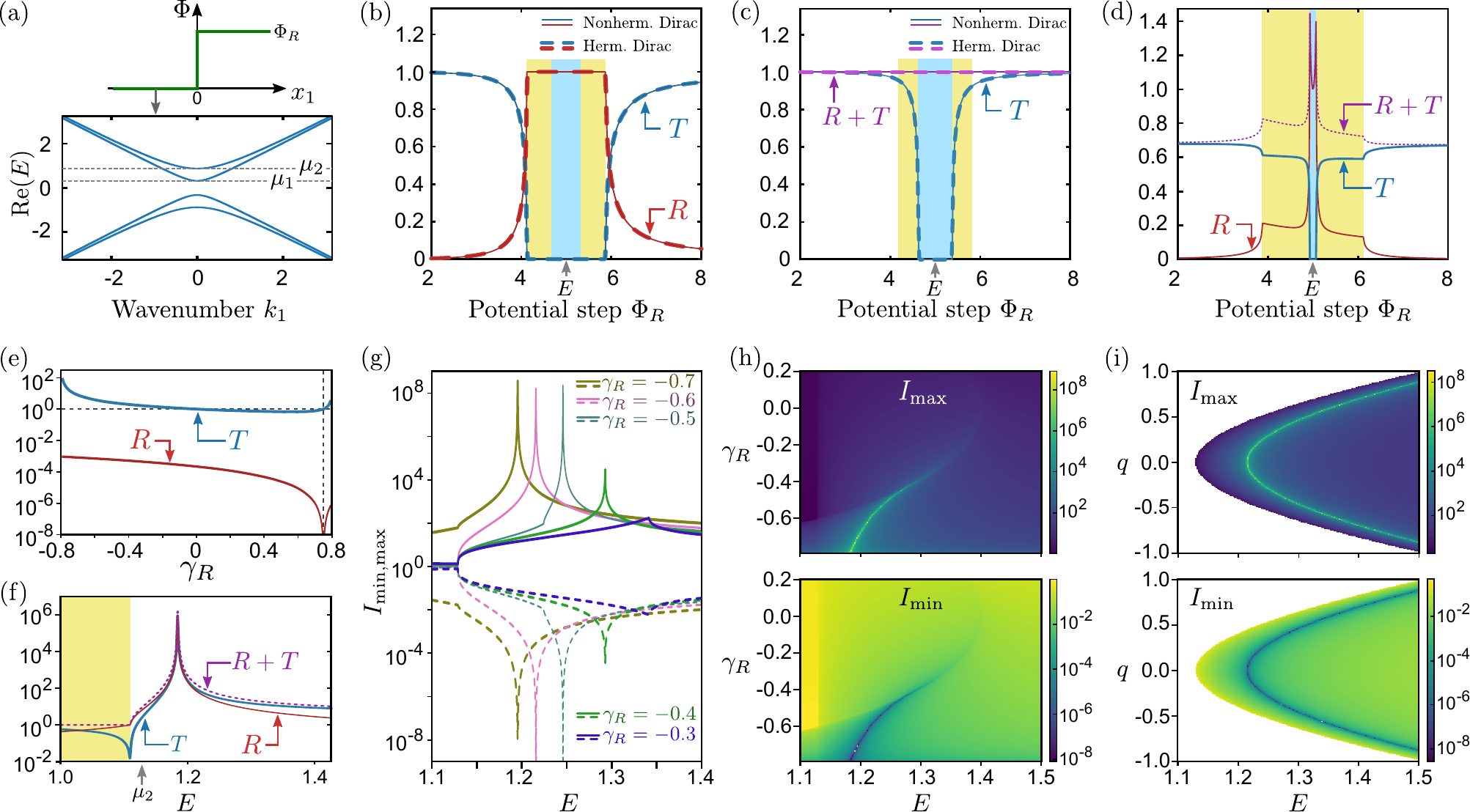}
\caption{(a) Setup for scattering calculations.  Top: Potential $\Phi(x_1)$ versus $x_1$.  Bottom: dispersion relation at normal incidence in the left domain, with $m_L=0.6$, $\gamma_L = 0.75$, and $\lambda_L=0.8$.  Semi-Hermiticity is unbroken, with $\mu_1 = 0.322$ and $\mu_2 = 0.878$. (b) Reflectance ($R$, red) and transmittance ($T$, blue) for a normally incident wave of $E = 5$ in mode $i = 2$, versus step height $\Phi_R$.  Both domains have the same $\mathcal{M}$.  Blue and yellow shaded regions respectively indicate the $n=1$ and $n=2$ gaps.  Results from the NHDE (solid lines) agree with the Hermitian Dirac equation (dashes). (c) Plot of $R$ and $R+T$ versus $\Phi_R$, with $\gamma_R = 0.6$ and $\lambda_R = 0.64$, and the incident wave in mode $i = 1$.  All other parameters are the same as in (b).  The domains have different $\gamma$ and $\lambda$, but the same $\gamma/\lambda$.  The NHDE (solid lines) again agrees with the Hermitian Dirac equation (dashes).  (d)  Plot of $R$, $T$, and $R+T$ versus $\Phi_R$, with $\lambda_R = 0.8$ and all other parameters the same as in (c).  The domains have unbroken semi-Hermiticity but different $\gamma/\lambda$.  Flux is not conserved, leading to anomalous tunneling and reflectance.  (e) Plot of $R$ and $T$ versus $\gamma_R$ at $E = 100$.  All other parameters are the same as in (d).  The vertical dashes mark the point $\gamma_R = \gamma_L \lambda_R/\lambda_L$.  (f) Plot of $R$, $T$, and $R+T$ versus $E$, for two semi-infinite domains with $m = 0.6$, $\lambda = 0.8$, and $\Phi = 0$.  The left and right domains have $\gamma_L = 0.6$ and $\gamma_R = -0.6$; semi-Hermiticity is unbroken.  The wave is incident from the left with $q = 0$.  $R$ and $T$ diverge at $E\approx 1.215$.  (g) Plot of $I_{\mathrm{max}}$ (solid lines) and $I_{\mathrm{min}}$ (dashes) versus $E$, for various $\gamma_R$.  $I_{\mathrm{max}}$ and $I_{\mathrm{min}}$ are the maximum and minimum eigenvalues of $S^\dagger S$, where $S$ is the scattering matrix.  When $I_{\mathrm{max}} \rightarrow \infty$ and $I_{\mathrm{min}} \rightarrow 0$, the domain wall is simultaneously a laser and coherent perfect absorber.  (h) Heat maps of $I_{\mathrm{max}}$ and $I_{\mathrm{min}}$ versus $E$ and $\gamma_R$.  (i) Heat maps of $I_{\mathrm{max}}$ and $I_{\mathrm{min}}$ versus $E$ and the transverse momentum $q$.  The region where the energy lies inside a gap, $E^2 < \mu_2^2 + q^2$, is left uncolored.  In (g)--(i), all other unspecified parameters are the same as in (f).}
\label{Fig:klein}
\end{figure*}

Let us consider how the NHDE behaves under such circumstances.  Starting from Eq.~\eqref{diracform2}, take an energy eigenstate
\begin{align}
  \left[\mathcal{M} -i \left(\Sigma_1\, \partial_1 + \Sigma_2\, \partial_2\right)+\Phi(x_1) \right] \psi=E \psi,
\label{eq:Kl1}
\end{align}
where $E$ is the incident energy, and $\Phi$ is a scalar potential that forms a step function along $x_1$:
\begin{equation}
  \Phi = \begin{cases}
    \Phi_L = 0, & x_1 < 0 \\ \Phi_R, & x_1 > 0.
  \end{cases}
  \label{potstep}
\end{equation}
This is plotted in the upper panel of Fig.~\ref{Fig:klein}(a).  In the following, we will also use $L$ and $R$ to label the other model parameters for the left and right domains.

We assume semi-Hermiticity is unbroken, so that the NHDE in each domain decomposes into two Hermitian Dirac equations with masses $\mu_1$ and $\mu_2$, as shown in the bottom panel of Fig.~\ref{Fig:klein}(a).  Let $q$ denote the conserved wavenumber in the $x_2$ direction.  In each spatial domain, the wavefunction can be decomposed into plane waves; the wavevectors are
\begin{align}
  \mathbf{k}_{n,\pm}^{d} &=\begin{bmatrix}
  s_\pm k_{n}^{d}\\ q
  \end{bmatrix}, \\
  k_{n}^{d} &= \sqrt{\left(E-\Phi_{d}\right)^2-\mu_n^2-q^2},
  \label{eq:Kl2}
\end{align}
where $d \in \{L,R\}$, and $s_\pm \in \{+, -\}$ is chosen so that the probability fluxes for $\mathbf{k}_{n+}^{d}$ and $\mathbf{k}_{n-}^{d}$ are respectively positive and negative along the $x_1$ direction (see Appendix~\ref{secProbability flux}).  These wavevectors are associated with the eigenequations
\begin{align}
  \left(\mathcal{M} + s_\pm \Sigma_1 k_{n}^{d} + \Sigma_2 q \right) u_{nd}^{\pm}
  &= \left(E -\Phi_{d}\right)u_{nd}^{\pm}, \\
  \mathcal{M}^2\, u_{nd}^{\pm} &= \mu_n^2 \, u_{nd}^{\pm}.
\end{align}
We seek a solution of the form
\begin{align*}
    \psi = e^{iqx_2} \left\{
    \begin{aligned}
    &u_{iL}^+ \, e^{ik_{i}^L x_1}
+ \sum_{n} r_n u_{nL}^- \; e^{-ik_{n}^L x_1}, &&x_1<0, \\
    &\sum_{n} t_{n} {u}_{nR}^+ \; e^{ik_{n}^R x_1}, &&x_1>0,
    \end{aligned}
    \right.
\end{align*}  
consisting of an incident eigenstate $u_{i+}^L$ ($i\in\{1,2\}$), and reflected and transmitted waves with coefficients $\{r_n\}$ and $\{t_n\}$ respectively.  These can be calculated by using the condition that the wavefunction is continuous at $x_1 = 0$ (see Appendix~\ref{secProbability flux}).
   
In Fig.~\ref{Fig:klein}(b), we plot the reflectance $R = \sum_n|r_n|^2$ (solid red lines) and transmittance $T = \sum_n|t_n|^2$ (solid blue lines) as a function of the potential step height $\Phi_R$.  The incident wave is prepared with $i = 2$, $q=0$ (normal incidence), and fixed $E$ (indicated by an arrow on the horizontal axis).  When the incident energy $E$ is detuned from the $n = 2$ gap, we observe Klein tunneling: $T \rightarrow 1$ and $R \rightarrow 0$.  When $E$ lies in the gap, $T \rightarrow 0$ and $R\rightarrow 1$ as expected; note that this includes a range where $E$ is outside the $n = 1$ gap, consistent with the expectation that the two Dirac subsystems are decoupled.  Throughout the plot, flux is conserved ($R+T=1$).  For comparison, the red and blue dashes show the reflectance and transmittance for a 1D Hermitian Dirac Hamiltonian $\mu \sigma_3 - i \sigma_1  \, \partial_1$, using the Dirac mass $\mu_1$.  Evidently, the non-Hermitian model and the Hermitian Dirac equation give the same results for all values of the potential step $\Phi_R$.  The results are also in exact agreement when the incident wave is prepared with $i = 1$, as well as for oblique incidence ($q \ne 0$).

Next, we consider what happens when the right domain not only has a different scalar potential, but also has different $\mathcal{M}$ parameters.  Fig.~\ref{Fig:klein}(c) shows the case where $m_R =0.6$, $\gamma_R = 0.6$, and $\lambda_R = 0.64$, with the left domain kept the same as before.  Both domains have unbroken semi-Hermiticity, but different Dirac masses: $\mu_{1L} = 0.322$, $\mu_{2L} = 0.878$, $\mu_{1R} = 0.377$, and $\mu_{2R} = 0.823$.  However, $\gamma_L/\lambda_L = \gamma_R/\lambda_R$, so according to Eq.~\eqref{eq:phin} they have the same decomposition into Dirac sub-Hamiltonians.  We find that the NHDE yields exactly the same results (solid lines) as the corresponding Hermitian Dirac equation \eqref{2dherm} (dashes).  In particular, flux is conserved, and there is no coupling between the two Dirac subsystems.

In Fig.~\ref{Fig:klein}(d), we investigate the situation where $\gamma/\lambda$ is not the same in both domains.  The left domain and the incident wave are the same as in the previous subplot, while the right domain has $m_R=0.6$, $\gamma_R = 0.6$, and $\lambda_R = 0.8$, corresponding to unbroken semi-Hermiticity with the Dirac masses $\mu_{1R} = 0.07$ and $\mu_{2R} = 1.13$ (the gaps are shaded in yellow and blue).  For $E \sim \Phi_R$, the value of $R$ can be significantly above unity.  On the other hand, when $E$ is far detuned from the gaps of the right domain, the system exhibits behavior similar to but distinct from Klein tunneling: $R \approx 0$, but $T < 1$ (thus, flux is not conserved).

Fig.~\ref{Fig:klein}(e) shows $R$ and $T$ versus $\gamma_R$, for $E = 100$ (far detuned from the Dirac mass gap), with all other parameters including $\gamma_L$ kept the same as in Fig.~\ref{Fig:klein}(d).  For different $\gamma_R$, either $T < 1$ or $T > 1$ can be achieved.  For $\gamma_R/\lambda_R = \gamma_L /\lambda_L$ (marked by vertical dashes), the system exhibits true Klein tunneling with $R \rightarrow 0$ and $T \rightarrow 1$, similar to Fig.~\ref{Fig:klein}(c).  However, even away from this point, $R$ remains small.



A more dramatic example of flux nonconservation induced by a domain wall can be seen in Fig.~\ref{Fig:klein}(f).  Here, the two domains have the same scalar potential $\Phi$, and the same $\mathcal{M}$ parameters except for $\gamma$, which switches sign across the domain wall.  Both $R$ and $T$ diverge at a certain value of $E$, indicating the onset of lasing \cite{chong2010, chong2011pt}.

To further study this phenomenon, Fig.~\ref{Fig:klein}(g) shows the $E$-dependence of $I_{\mathrm{max}}$ and $I_{\mathrm{min}}$, defined respectively as the maximum and minimum eigenvalues of $S^\dagger S$, where $S$ is the scattering matrix for the domain wall (see Appendix~\ref{secProbability flux}).  $I_{\mathrm{max}}$ and $I_{\mathrm{min}}$ respectively represent the largest and smallest possible total output powers scattered from the domain wall, given one unit of total input power \cite{chong2011}.  For a Hermitian system, $S$ is unitary so $I_{\mathrm{max}} = I_{\mathrm{min}} = 1$, whereas a lasing threshold corresponds to $I_{\mathrm{max}} \rightarrow \infty$ (since a laser emits nonzero power even when there are no coherent input waves).  We find that each threshold laser solution for the domain wall is accompanied at the same $E$ by a zero of $I_{\mathrm{min}}$, corresponding to a coherent perfect absorber (CPA) solution \cite{chong2010, baranov2017}.  The domain wall thus functions as a CPA-laser: a scatterer that is poised at a lasing threshold, but can also perfectly absorb a specific set of input waves \cite{longhi2010cpalaser, chong2011pt}.  In Fig.~\ref{Fig:klein}(h)--(i), we show how the energy of the CPA-laser solutions varies with $\gamma_R$ and the transverse wavenumber $q$.  The CPA-laser solutions disappear for $\gamma_R \gtrsim -0.4$, as can also be seen in Fig.~\ref{Fig:klein}(g).

It is worth emphasizing that \textit{the domain wall itself} is acting here as a source or sink of flux.  Wave amplification and damping does not occur within the bulk domains, as these have unbroken semi-Hermiticity and host only real-energy plane wave solutions.  Note that the possibility of a thin-layer CPA (not a CPA-laser) has previously been explored \cite{tischler2006, Liu2014}.

\begin{figure}
\centering
\includegraphics[width=0.475\textwidth]{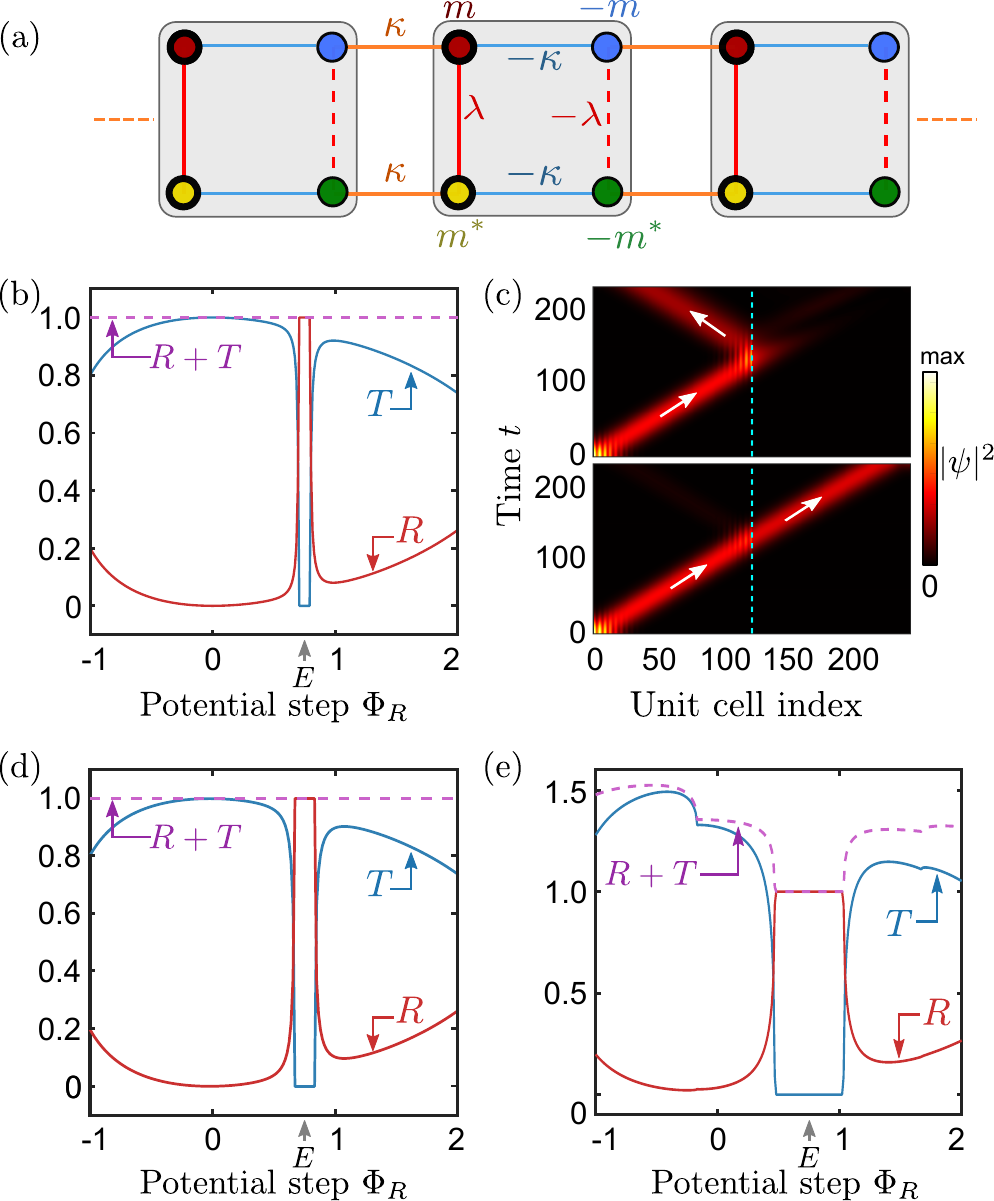}
\caption{(a) Schematic of the 1D lattice. (b) Plot of $R$, $T$, and $R+T$ versus $\Phi_R$ for a domain wall in the 1D lattice, calculated with a frequency-domain solver \cite{kwant} for $E=0.75$.  The lattice parameters are $m=0.6$, $\gamma = 0.58$, $\lambda=0.8$, and $\kappa=1$ for both domains. (c) Time-domain plots of $|\psi|^2$ for a 1D lattice of length 240 unit cells, with the wavefunction at $t = 0$ initialized to a right-moving wavepacket with central energy $E = 0.75$ and spatial width $\Delta x = 32$.  For $\Phi_R = 0.75 = E$ (upper panel), the wavepacket is almost entirely reflected at the potential step (cyan dashes), whereas for $\Phi_R = 0.95$ (lower panel), strong transmission occurs, consistent with the frequency-domain results.  All other parameters are the same as in (b).  (d) Similar to (b), but with $m_R=0.6$, $\gamma_R = 0.5437$, and $\lambda_R=0.75$.  Note that $\gamma_L/\lambda_L = \gamma_R/\lambda_R$ and $R + T = 1$.  (e) Similar to (d), but with $\lambda_R = 0.63$; now $\gamma_L/\lambda_L \ne \gamma_R/\lambda_R$, and flux conservation is violated. }
\label{Fig:klein_lattice}
\end{figure}

The above reflection and transmission behaviors can be reproduced in lattice models.  For example, Fig.~\ref{Fig:klein_lattice} shows results for $q = 0$ using a 1D lattice.  The design of the lattice, shown in Fig.~\ref{Fig:klein_lattice}(a), is a simplified version of the 2D lattice presented in Sec.~\ref{sec:2Dlattice} obtained by dimensional reduction to eliminate the $x_1$ direction by setting $k_1 = 0$.  The resulting 1D lattice model contains only real and reciprocal hoppings $\{\pm\kappa, \pm \lambda\}$.  In Fig.~\ref{Fig:klein_lattice}(b), we plot $R$, $T$, and $R+T$ for two semi-infinite lattice domains with the same parameters, aside from a potential step $\Phi_R$.  The behavior is similar to the continuum case shown in Fig.~\ref{Fig:klein}(b), except that the Klein tunneling breaks down when $E$ and $\Phi_R$ are too far detuned, due to the finite band-width of the lattice model.  Notably, $R + T = 1$ throughout the plot.  Time-domain simulations, plotted in Fig.~\ref{Fig:klein}(c), reveal the same behavior, i.e.~strong reflection when $E \approx \Phi_R$ and strong transmission when $E$ and $\Phi_R$ are detuned (but not too far detuned).

In Fig.~\ref{Fig:klein_lattice}(d), we show that the Hermitian Dirac-like behavior holds even when the lattice parameters in the two domains are different (in this case, $\gamma_L \ne \gamma_R$ and $\lambda_L \ne \lambda-R$), provided that $\gamma_L/\lambda_L = \gamma_R/\lambda_R$.  Similar to the continuum case shown in Fig.~\ref{Fig:klein}(c), we find that $R + T = 1$.  On the other hand, we see in Fig.~\ref{Fig:klein_lattice}(e) that when $\gamma_L/\lambda_L \ne \gamma_R/\lambda_R$, flux conservation breaks down, similar to Fig.~\ref{Fig:klein}(d).  In all of these plots, the parameters in both domains are chosen such that the infinite-lattice bandstructure has unbroken semi-Hermiticity throughout the 1D Brillouin zone.

\begin{figure*}
  \centering
  \includegraphics[width=0.99\textwidth]{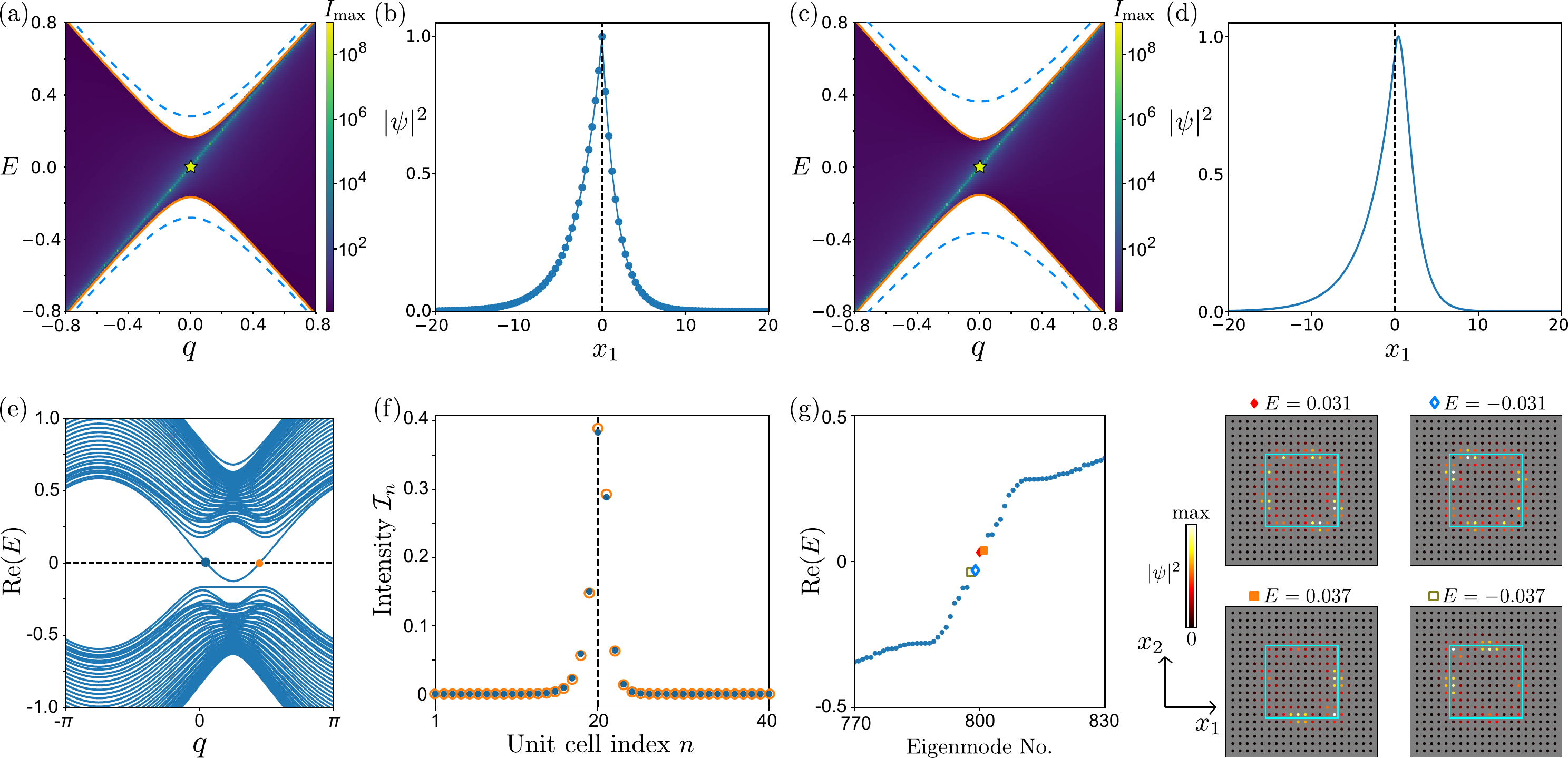}
  \caption{(a) Dispersion relation of boundary states.  The heat map of $I_{\mathrm{max}}$, the largest eigenvalue of $S^\dagger S$, is plotted versus real $E$ and $q = k_2$.  A domain wall at $x_1 = 0$ separates semi-infinite domains with $m_{L}=0.7$, $\gamma_L=0.5$, $\lambda_L=1.0$, $m_R=0.8$, $\gamma_R=0.3$, and $\lambda_R=0.6$.  Note that $\gamma_L/\lambda_L = \gamma_R/\lambda_R$.  The plotted $E$ values lie in the mass gaps, so $I_{\mathrm{max}} \rightarrow \infty$ indicates a boundary state (see Appendix~\ref{secProbability flux}).  Solid orange (dashed blue) lines indicate band edges of the $L$ ($R$) domain; some band edges lie outside this plot.  (b) Intensity profile of the state marked by a star in (a), with $E = 0$.  Results from the NHDE (solid lines) match the Hermitian Dirac edge state (dots).  (c) Similar to (a), but with $m_{L}=m_R=0.8$, $\gamma_L=0.3$, $\gamma_R = 0.9$, and $\lambda_L=\lambda_R=1.0$.  In this case, $\gamma_L/\lambda_L \ne \gamma_R/\lambda_R$. (d) Intensity profile of the $E = 0$ state marked by a star in (c).  (e) Bandstructure for a 2D lattice based on Fig.~\ref{fig:model}(a), which forms a strip 40 unit cells wide along $x_1$ and periodic along $x_2$.  A domain wall runs parallel to $x_2$, between unit cells 20 and 21, separating domains with the same $\{m,\gamma,\lambda\}$ as in (a), and $\kappa_L = \kappa_R = 0.4$.  All energy eigenvalues are real.  (f) Intensity profiles of two boundary states marked by dots in (e), at $E = 0$ with $q = 0.142$ and $q = 1.405$ (orange and blue dots).  For each unit cell $n$, the mean intensity on its four sites is plotted.  (g) Left: energy diagram for a finite $20\times20$ lattice consisting of a $10\times10$ inner region with the parameters of the $L$ lattice in (e), and an outer region with the parameters of the $R$ lattice in (f).  All eigenenergies are real. Right: intensity profiles for the four eigenstates closest to $E = 0$.  The domain wall is marked in cyan. }
    \label{fig:Domainwall}
\end{figure*}

\section{Boundary states}
\label{sec:boundstates}

The Hermitian Dirac equation supports boundary solutions that are exponentially localized to domain walls.  These play a particularly important role in the theory of band topology \cite{hasan2010colloquium, Cayssol2021}.  For example, the Su-Schrieffer-Heeger (SSH) lattice model reduces to a 1D Dirac Hamiltonian in the limit of long wavelengths and a small band gap.  The SSH model is known to host zero-energy boundary states tied to a topological invariant, the Zak phase \cite{SSH1979, Zak1989}, and in the continuum limit these states reduce to the Jackiw-Rebbi zero mode for a 1D Dirac equation with a mass inversion \cite{JackiwRebbi1976}.  In a similar fashion, the 2D Haldane model can be reduced to a pair of 2D Dirac Hamiltonians, and its topological phase transitions can be understood in terms of the sign changes of the two Dirac masses \cite{haldane1988model}.

In this section, we show that the 2D NHDE can host boundary states with real energies, not only when $\gamma/\lambda$ is constant (i.e., there is a consistent decomposition into Hermitian Dirac subsystems), but also when it is not.  The latter case is notable because, in Sec.~\ref{sec:klein}, we saw that such a domain wall generally violates flux conservation.

As with Sec.~\ref{sec:klein}, we assume a setup with a domain wall at $x_1 = 0$, with the parameters in the left ($L$) and ($R$) domains given by $m_{L/R}$, $\gamma_{L/R}$, etc.  Both domains have $\Phi = 0$ and unbroken semi-Hermiticity.  We look for boundary states with real $E$ lying in the mass gaps, so that the wavefunction decays exponentially away from the domain wall in both directions.  As described in Appendix~\ref{secProbability flux}, the scattering matrix framework used for studying reflection and transmission in Sec.~\ref{sec:klein} can be adapted for finding boundary states.  When $E$ lies in the mass gaps, we can pick an appropriate mode-sorting convention so that boundary states correspond to poles of the scattering matrix $S$.

Fig.~\ref{fig:Domainwall}(a) shows the dispersion relation for the boundary states at a domain wall.  The colors represent the heat map of $I_{\mathrm{max}}$, defined as before as the maximum eigenvalue of $S^\dagger S$, plotted against $E$ (taken to be real) and $q$ (the conserved momentum along $x_2$, parallel to the domain wall).  The domains have Dirac masses $\mu_{1L} = 1.57$, $\mu_{2L}=-0.17$, $\mu_{1R} = 1.32$, and $\mu_{2R} = 0.28$.  We see that there is a branch of real-energy boundary states ($I_{\mathrm{max}} \rightarrow \infty$) within the Dirac mass gap, with a chiral dispersion relation.  In this plot, $\gamma_L/\lambda_L = \gamma_R/\lambda_R$, so the boundary states can be associated with a band inversion in one of the Dirac subsystems, $\mu_{2L} < 0$ and $\mu_{2R} > 0$.  In Fig.~\ref{fig:Domainwall}(b), the intensity profile of the boundary state at $q = 0$ is plotted as a solid curve.  The intensity profile for the Hermitian Dirac boundary state, derived from Eq.~\eqref{2dherm}, is plotted using dots and exactly matches the NHDE result.  (Note that the decay length is shorter in the right domain, as $\mu_{2R} > \mu_{2L}$.)

In Fig.~\ref{fig:Domainwall}(c), we repeat the calculation for a different set of parameters.  In this case, $\gamma_L/\lambda_L \ne \gamma_R/\lambda_R$ so the two domains do not have the same decomposition into Hermitian Dirac subsystems.  Nonetheless, we still find a family of boundary states with real energies and chiral dispersion.  The intensity profile for the $q = 0$ boundary state is plotted in Fig.~\ref{fig:Domainwall}(d).  Note that the peak occurs slightly to the right of the domain wall (i.e., it is not monotonically decreasing throughout $x_1 > 0$).  This is consistent with expectations: the wavefunction is no longer a solution to a single decoupled Hermitian Dirac equation with a single-exponential $x_1$-dependence, and the solution in each domain is a superposition of two exponentials drawn from both Dirac subsystems.

Note that the real-energy boundary states in Fig.~\ref{fig:Domainwall}(c)--(d) are purely induced by gain/loss.   The parameters in the two domains differ only in the value of $\gamma$, a local gain/loss term appearing in the imaginary diagonal components of the Hamiltonian [Eq.~\eqref{wdef}].  In recent years, gain/loss-induced boundary states have been observed in a number of other non-Hermitian models \cite{TakataNotomi2018, Kawabata2018, Luo2019, Wu2020}.  However, those earlier models have exhibited complex bulk bandstructures, meaning that flux is not conserved either in the bulk nor at the domain wall; in some cases, the boundary states themselves also have $\mathrm{Im}(E) \ne 0$.  Our results here demonstrate that gain/loss-induced boundary states with $\mathrm{Im}(E) = 0$ can be achieved in a non-Hermitian system whose bulk states all have real energy.  This is the case even if the domain wall violates flux conservation for reflection/transmission outside the mass gap (as we saw in Sec.~\ref{sec:klein}).

Similar boundary states also occur in the 2D lattice model we introduced in Section \ref{sec:2Dlattice}.  Take a strip that is 40 unit cells wide along $x_1$, and periodic along $x_2$.  Let a domain wall run parallel to the strip, with different lattice parameters on each side; we set $m$, $\gamma$, and $\lambda$ to the same values as in Fig.~\ref{fig:Domainwall}(a), so that there is a mass inversion across the domain wall, with the same $\kappa$ in both domains.  The resulting band diagram is shown in Fig.~\ref{fig:Domainwall}(e).  The spectrum is entirely real, and features a bulk gap around $E = 0$; this gap is sizeable compared to the band-width, and the band edges are non-hyperbolic. The gap is spanned by boundary states (there are both positive- and negative-velocity branches since the lattice model contains two Dirac points, as mentioned in Sec.~\ref{sec:2Dlattice}; this is similar to the case of a valley Hall lattice \cite{RevModPhys.91.015006}). As shown in Fig.~\ref{fig:Domainwall}(f), these states are exponentially localized to the boundary.

In Fig.~\ref{fig:Domainwall}(g), we consider the case of a finite 2D lattice sample, consisting of $20\times20$ unit cells with an inner $10\times 10$ region, and open (Dirichlet-like) external boundaries.  The inner (outer) domain is assigned the same parameters as the $L$ ($R$) domain in the preceding strip calculation.  The spectrum is again observed to be entirely real, and is plotted in the left panel.  In the right panels, we plot the intensity profiles for the four states with energies closest to $E = 0$.  These are observed to be strongly localized to the square-shaped domain wall.

\section{Conclusion}

We have presented a (2+1)-dimensional non-Hermitian Dirac equation (NHDE) that exhibits numerous similarities to the Hermitian Dirac equation, but also important differences.  The model relies on the non-Hermitian symmetries introduced in Ref.~\onlinecite{xue2020non}, referred to in this paper as ``semi-Hermiticity'', which allow a non-Hermitian Hamiltonian to have real spectra with pairwise-orthogonal eigenstates.  In the context of the NHDE, under translationally invariant conditions, we used semi-Hermiticity to derive an explicit relationship between the plane wave eigenstates of the NHDE and a pair of Hermitian Dirac equations.

We then considered two uniform domains with different scalar potentials and/or NHDE parameters, with each having unbroken semi-Hermiticity so that their bulk solutions are like Hermitian Dirac modes.  Remarkably, depending on the parameter choices, reflection and transmission at the interface can be identical to the Hermitian case (incuding Klein tunneling \cite{Klein1929}), or significantly different due to the breakdown of flux conservation.  In the latter case, the flux nonconservation occurs at the domain wall, not in the bulk, and can be strong enough for the domain wall to act as a laser and coherent perfect absorber \cite{chong2010, baranov2017, longhi2010cpalaser, chong2011pt}.  These domain walls also host exponentially localized boundary states with real energies.

In future work, it would be interesting to experimentally realize some of the above phenomena, such as the domain wall laser/absorber, using appropriately-designed classical wave metamaterials.  It would also be worthwhile to investigate other phenomena associated with the Hermitian Dirac equation, such as zitterbewegung \cite{ZhangXd2008, Longhi2010zitter, Dreisow2010} and Dirac solitons \cite{fushchych2006, Poddubny2018}, in the context of the NHDE.  In particular, whether relativistic symmetries can play any meaningful role in this non-Hermitian system is presently unclear.

\appendix

\section{Scattering Calculations}
\label{secProbability flux}

Take the NHDE \eqref{diracform2}, with all model parameters constant in space and time.  As shown in Sec.~\ref{sec:basics}, when semi-Hermiticity is unbroken, there can be plane wave solutions $\psi(\mathbf{r}, t) = u \exp[i(\mathbf{k}\cdot\mathbf{r}- Et)]$, where $\mathbf{k}\in\mathbb{R}^2$, $E = \pm \sqrt{\mu^2 + |\mathbf{k}|^2} \in \mathbb{R}$ where $\mu \in \mathbb{R}$ is an eigenvalue of $\mathcal{M}$, and $u$ satisfies
\begin{align}
  \left(\mathcal{M} + \Sigma_1\, k_1 + \Sigma_2\, k_2\right) u = E u.
  \label{eq:Mu}
\end{align}
The probability flux $\mathbf{J} = (J_1, J_2)$ carried by these plane waves can be determined by taking
\begin{align*}
  \psi^\dagger & \frac{\partial \psi}{\partial t} + \frac{\partial \psi^\dagger}{\partial t} \psi \\
  &=-i \psi^\dagger \left(\mathcal{M} - \mathcal{M}^\dagger\right) \psi - \partial_1  \left( \psi^\dagger \Sigma_1 \psi \right)- \partial_2  \left( \psi^\dagger \Sigma_2 \psi \right).
\end{align*}
Comparing this to
\begin{align*}
\frac{\partial}{\partial t} \left(\psi^\dagger \psi\right) = [\text{gain/loss term}] - \nabla \cdot J
\end{align*}
yields the result
\begin{align}
J_j =  u^\dagger \Sigma_j u.
\label{eq:APF4}
\end{align}
This derivation is unaltered if we introduce a uniform scalar potential $\Phi$.

Now consider a domain wall at $x_1 = 0$, separating two spatially uniform domains indexed by $d = L$ (i.e., $x_1 < 0$) and $d = R$ (i.e., $x_1 > 0$).  The domains may have different $\mathcal{M}$ and scalar potential $\Phi$.  Suppose that at a given energy $E \in \mathbb{R}$, each domain has four real eigenvalues $\{\pm \mu_{1d}, \pm \mu_{2d}\}$---i.e., $E$ lies outside the Dirac mass gaps.  Then each domain hosts plane waves with eigenvectors
\begin{align}
  \{ u_{nd}^s \},  \;\;\mathrm{where}\;s = \pm,\;\;
  n \in \{1,2\}, \;\;
  d \in \{L/R\}.
  \label{eq:udef}
\end{align}
Each $u_{nd}^s$ is associated with a wave-vector $(\pm k_{nd}, q)$, with the sign of the first component chosen so that the direction of probability flux is $s$.  We use Eq.~\eqref{eq:APF4} to impose the normalization $|J_1| = 1$.  The eigenvectors satisfy Eq.~\eqref{eq:Mu} along with
\begin{align}
  \mathcal{M}^2 u_{nd}^s &= \mu_n u_{nd}^s, \\
  E^2 &= \mu_n^2 + k_{n,d}^2 + q^2.
\end{align}
In each domain $d$, we can construct the superposition
\begin{align*}
  \psi(x_1) = \sum_{sn} a^s_{nd} u_{nd}^s e^{ \pm i k_{nd} x_1 }.
\end{align*}
Continuity of the wavefunction at $x_1 = 0$ then implies
\begin{multline}
  \begin{bmatrix}
    u_{1L}^+ & u_{2L}^+ & -u_{1R}^- & -u_{2R}^-
  \end{bmatrix}_j \mathbf{a}_{\mathrm{in}}\\
  = \begin{bmatrix}
    -u_{1L}^- & -u_{2L}^- & u_{1R}^+ & u_{1R}^+
  \end{bmatrix}_j \mathbf{a}_{\mathrm{out}},
\end{multline}
for each eigenvector component $j = 1,\dots,4$, where
\begin{equation}
  \mathbf{a}_{\mathrm{in}} = \begin{bmatrix}
      a^+_{1L} \\ a^+_{2L} \\ a^-_{1R} \\ a^-_{2R}
    \end{bmatrix}, \;\;\;
    \mathbf{a}_{\mathrm{out}} =
    \begin{bmatrix}
    a^-_{1L} \\ a^-_{2L} \\ a^+_{1R} \\ a^+_{2R}
  \end{bmatrix}.
\end{equation}
We use this define the scattering matrix $S$:
\begin{equation}
  S(E,q)\; \mathbf{a}_{\mathrm{in}} = \mathbf{a}_{\mathrm{out}}.
\end{equation}
From this, we can also obtain the reflectance ($R$) and transmittance ($T$) results discussed in Sec.~\ref{sec:klein}.  For example, for a wave incident from the $L$ domain in mode $n = 1$, we set $\mathbf{a}_{\mathrm{in}} = [1,0,0,0]^T$, and from $\mathbf{a}_{\mathrm{out}}$ we obtain $R = \sum_n |a_{nL}^-|^2$ and $T = \sum_n |a_{nR}^+|^2$.

We can seamlessly handle the case where $E$ lies inside a Dirac mass gap, by adapting \eqref{eq:udef} so that $s = +$ ($s = - $) represents an evanescent wave decaying toward the right (left).  With this convention, setting the associated component of $\mathbf{a}_{\mathrm{in}}$ to zero is equivalent to excluding ``unphysical'' waves that grow exponentially away from the domain wall.  We do not use use Eq.~\eqref{eq:APF4} to normalize evanescent waves; their associated components in $\mathbf{a}_{\mathrm{out}}$ (which correspond to steady-state modes localized at the domain wall) are excluded when calculating $R$, $T$, $I_{\mathrm{min}}$, and $I_{\mathrm{max}}$ in Sec.~\ref{sec:klein}.

\input{nh-dirac.bbl}
\end{document}

%% file: nh-dirac.bbl
%